\def\bnslash{\bar n\!\!\!\slash}

\def\OMIT#1{}

\newcommand{\nn}{\nonumber}

\newcommand{\bn}{{\bar n}}
\newcommand{\bea}{\begin{eqnarray}}
\newcommand{\eea}{\end{eqnarray}}

\newcommand{\be}{\begin{equation}}
\newcommand{\ee}{\end{equation}}

\documentclass{ws-ijmpcs}

\begin{document}

\markboth{Mantry and Petriello}
{Transverse Momentum Distributions from Effective Field Theory}

%
\catchline{}{}{}{}{}
%

\title{TRANSVERSE MOMENTUM DISTRIBUTIONS FROM EFFECTIVE FIELD THEORY}

\author{Sonny Mantry\footnote{Invited Talk presented at the workshop on ``QCD Evolution of Parton Distributions: from Collinear to Non-collinear Case", April 8th-9th, 2011, Thomas Jefferson National Accelerator Facility, Newport News, VA, USA.}}

\address{University of Wisconsin at Madison\\
Madison, WI  53706, USA\\
mantry147@gmail.com}

\author{Frank Petriello}

\address{High Energy Physics Division, Argonne National Laboratory \\
Argonne, IL 60439, USA\\}
\address{Department of Physics \& Astronomy, Northwestern University\\
Evanston, IL 60208, USA\\
f-petriello@northwestern.edu}

\maketitle

\begin{abstract}

We review a new approach to calculating transverse momentum distributions of the Higgs and electroweak gauge bosons using the Soft-Collinear Effective Theory.  We derive a factorization theorem for transverse momentum distributions in terms of newly-defined impact-parameter beam functions (iBFs) and an inverse soft function (iSF).  The iBFs correspond to completely unintegrated parton distribution functions and provide interesting probes of momentum distributions within nucleons.  The numerical matching between the low and high transverse momentum regions is improved in this approach with respect to standard techniques.  We present results for next-to-leading logarithmic resummation for the Higgs and Z-boson distributions and give a comparison with Tevatron data.

\end{abstract}

\section{Introduction}	

Low transverse momentum ($p_T$) distributions of the Higgs and electroweak
gauge bosons play an important role in numerous physics studies, including Higgs searches, precision measurements
of the $W$-boson mass, tests of perturbative QCD, and in probing transverse momentum
dynamics in the nucleon. In the region $p_T\ll M$, where $M$ denotes the mass of the Higgs
or electroweak boson in question, large double logarithms  in $p_T/M$ spoil the convergence
of the perturbative expansion in the strong coupling. A proper description of the transverse
momentum distributions in this region requires resummation of these large logarithms. Extensive
studies~[\refcite{Dokshitzer:1978yd,Parisi:1979se,Curci:1979bg,Collins:1981uk,Collins:1984kg,Kauffman:1991jt,Yuan:1991we,Ellis:1997ii,Kulesza:2002rh,Berger:2002ut,Bozzi:2003jy,Bozzi:2005wk,Bozzi:2010xn,Gao:2005iu,Becher:2010tm,Idilbi:2005er}] of low $p_T$ resummation have been performed.  Most are based on the well-known Collins-Soper-Sterman (CSS) formalism.

More recently in Refs.~[\refcite{Mantry:2009qz,Mantry:2010mk,Mantry:2010bi}], low $p_T$ resummation was studied using the Soft-Collinear Effective Theory (SCET)~[\refcite{Bauer:2000yr,Bauer:2001yt,Bauer:2002nz}]. In the region $M\gg p_T \gg \Lambda_{QCD}$, the schematic form of the factorization theorem is
\begin{eqnarray}
\label{fac-1}
\frac{d^2\sigma}{dp_T^2 \> dY} \sim H\otimes {\cal G}\otimes f\otimes f.
\end{eqnarray}
Here, $H$ is a hard function that encodes the physics of the hard production vertex and whose renormalization group (RG) evolution sums the
logarithms of $p_T/M$.  ${\cal G}$ is
a gauge invariant Transverse Momentum Function (TMF) at the scale $\mu_T \sim p_T$ that describes the dynamics of
initial state soft and collinear  emissions   that recoil against the heavy boson. 
$f$ denotes the standard PDF and is evaluated at $\mu_T\sim p_T$ as determined by DGLAP evolution. In the region 
where $p_T\sim \Lambda_{QCD}$, the TMF  beomes non-perturbative and encodes the  transverse momentum
dynamics in the nucleons and of the soft radiation in the process.

A smooth transition between the regions of non-perturbative and perturbative $p_T$ is achieved by implementing
a model~\cite{Mantry:2010bi}  for the TMF ${\cal G}$ in the non-perturbative region such that it has the correct RG evolution properties
and smoothly reduces to the perturbative result for increasing $p_T$. The field-theoretic definition of ${\cal G}$ provides
further insight into  the structure of the factorization theorem. In particular, we  identify new objects called Impact-parameter
Beam Functions (iBFs) which correspond to fully unintegrated PDFs. The form of the factorization in terms of
these iBFs has differences compared to the more traditional TMDPDF formalism (see other proceedings~[\refcite{Collins:2011ca,Aybat:2011vb,Cherednikov:2011ku}] of this workshop for the most current overview). We give a brief discussion on this in the next section. In the following sections,
we also give an overview of the factorization theorem, numerical results, and provide comparison with data.

\section{Overview of Factorization Theorem}

The first step in the derivation of the factorization theorem is to integrate out the hard production scale by matching QCD
onto the SCET which describes the dynamics of initial state collinear and soft emissions
that recoil against the heavy boson. These modes in the SCET have the  momentum  scalings 
\begin{eqnarray}
p_n \sim M (\eta^2, 1, \eta), \>\>\>\> p_{\bar{n}} \sim M (1,\eta^2,\eta), \>\>\>\> p_{s}\sim M(\eta,\eta,\eta), \>\>\>\>\eta \sim\frac{p_T}{M},
\end{eqnarray}
corresponding to the $n$-collinear, $\bar{n}$-collinear, and soft modes respectively. The heavy boson
acquires finite transverse momentum by recoiling against these soft and collinear  modes which
have transverse momenta of order $p_T$. Using the soft-collinear decoupling property of the SCET,
the dynamics of the collinear and soft emissions are separated into two zero-bin subtracted~\cite{Manohar:2006nz,Idilbi:2007ff} iBFs ($B_n, B_{\bar{n}}$) and a soft function ($S$)  respectively.
The  zero-bin subtraction is necessary to avoid double counting the soft
region. Such a subtraction also appears in the TMDPDF formalism~\cite{Collins:1999dz}. Using the equivalence of the zero-bin and soft subtractions one can rewrite the factorization theorem in terms of iBFs ($\tilde{B}_n, \tilde{B}_{\bar{n}}$) defined without the soft
zero-bin subtractions and an Inverse Soft Function (iSF). In the rest of this article,  iBFs will refer to the $\tilde{B}_n, \tilde{B}_{\bar{n}}$ functions
defined without the soft zero-bin subtraction.
For perturbative values of the transverse momentum $p_T\gg \Lambda_{QCD}$, the iBFs are matched onto the standard
PDFs after performing an operator product expansion in $\Lambda_{QCD}/p_T$. The Wilson coefficients of the iBF to PDF
matching combined with the iSF give the TMF function ${\cal G}$. We schematically summarize these steps
in deriving the factorization theorem below:
\begin{eqnarray}
\label{fac-schem}
\frac{d^2 \sigma}{dp_T^2 dY} &\sim& \int PS \, |{\cal M}_{QCD}|^2 \\ \nonumber
&\downarrow& \text{(match QCD to }\text{SCET}_{p_T})\\ \nonumber
&\sim& \int PS \, |C \otimes \langle {\cal O}_{SCET} \rangle |^2 \\ \nonumber
&\downarrow& \text{(SCET soft-collinear decoupling)}\\ \nonumber
&\sim& H \otimes B_n \otimes  B_{\bar{n}} \otimes S \\ \nonumber
&\downarrow& \text{(zero-bin and soft subtraction equivalence)}\\ \nonumber
&\sim& H \otimes  \tilde{B}_n  \otimes \tilde{B}_{\bar{n}}  \otimes S^{-1} \\ \nonumber
&\downarrow& \text{( iBF to PDF matching)}\\ \nonumber
&\sim& H \otimes \underbrace{\left[{\cal I}_{n} \otimes {\cal I}_{\bar{n}} \otimes S^{-1}\right]}_{{\cal G}} \otimes f_i \otimes f_j,
\label{eq:fact_outline}
\end{eqnarray}
where the coefficients ${\cal I}_{n,\bar{n}}$ arise from the iBF-to-PDF matching
\begin{eqnarray}
\label{iBFPDF}
 \tilde{B}_n = {\cal I}_n \otimes f , \>\>\>\>   \tilde{B}_{\bar{n}} = {\cal I}_{\bar{n}} \otimes f.
\end{eqnarray}
For $p_T\gg \Lambda_{QCD}$, the iBFs, which correspond to fully unintegrated PDFs, describe the evolution of the initial state
nucleons followed by their disintegration into a jet of collinear radiation that recoils against the heavy boson.
Analogous beam functions are known to arise in other processes at hadron colliders~[\refcite{Fleming:2006cd,Stewart:2009yx}].
For $p_T\sim \Lambda_{QCD}$, a perturbative matching of the iBF onto the PDF is no longer
valid. In fact, in this region, the iBFs and the iSF are non-perturbative. Eq.(\ref{iBFPDF}) should no longer
be viewed as a perturbative matching but instead must be viewed as a definition of the new non-perturbative
functions ${\cal I}_n$ and ${\cal I}_{\bar{n}}$. Thus, the TMF function ${\cal G}$ is non-perturbative 
for $p_T \sim \Lambda_{QCD}$. It encodes the transverse momentum dynamics within the nucleon
and in the soft radiation that determines
the $p_T$ of the heavy boson. 

For the sake of brevity, we focus most discussion here specifically for the case of the Z-boson $p_T$ distribution. We refer the reader for  details
of the derivation,  notation, and  the case of Higgs boson distribution to Refs.~[\refcite{Mantry:2009qz,Mantry:2010mk}]. The factorization theorem for the Z-boson distribution  takes the form
\begin{eqnarray}
\label{intro-DY}
\frac{d^2\sigma}{dp_T^2\> dY}&=& \frac{\pi^2 }{N_c^2}   \int_0^1 dx_1 \int_0^1 dx_2\int_{x_1}^1 \frac{dx_1'}{x_1'} \int_{x_2}^1 \frac{dx_2'}{x_2'}    \\
&\times&   H_Z^{q}(x_1x_2Q^2,\mu_Q;\mu_T) \>{\cal G}^{qrs}(x_1,x_2,x_1',x_2',p_T,Y,\mu_T) f_r(x_1',\mu_T)  f_s(x_2',\mu_T),\nonumber
\end{eqnarray}
where the indices $r,s$ run over the initial partons. The TMF function ${\cal G}^{qrs}$ is given by
\begin{eqnarray}
\label{intro-DY-2}
 &&{\cal G}^{qrs}(x_1,x_2,x_1',x_2',p_T,Y,\mu_T)=  \int \frac{d^2b_\perp}{(2\pi)^2} J_0\big [b_\perp p_T\big ]\>\int dt_n^+ dt_{\bar{n}}^- \nn \\
 &\times& \> {\cal I}_{n;q r}(\frac{x_1}{x_1'}, t_n^+,b_\perp,\mu_T)\>{\cal I}_{\bar{n};\bar{q} s}(\frac{x_2}{x_2'}, t_{\bar{n}}^-,b_\perp,\mu_T)\\
&\times& {\cal S}^{-1}_{qq}(x_1 Q-e^{Y}\sqrt{\text{p}_T^2+M_Z^2}-\frac{t_{\bar{n}}^-}{x_2 Q}, x_2 Q-e^{-Y}\sqrt{\text{p}_T^2+M_Z^2}- \frac{t_n^+}{x_1 Q},b_\perp,\mu_T). \nn
\end{eqnarray}
The iSF $ {\cal S}^{-1}_{qq}(\tilde{\omega}_1,\tilde{\omega}_2,b_\perp,\mu)$ is given by
\begin{eqnarray}
 {\cal S}^{-1}_{qq}(\tilde{\omega}_1,\tilde{\omega}_2,b_\perp,\mu) &=& \int \frac{db^+db^-}{16\pi^2} e^{\frac{i}{2}\tilde{\omega}_1b^+} e^{\frac{i}{2}\tilde{\omega}_2b^-} S_{qq}^{-1}(b^+,b^-,b_\perp,\mu)\,
\end{eqnarray}
where the position-space soft function $S_{qq}(b^+,b^-,b_\perp,\mu)$ is given by the vacuum matrix element of soft Wilson lines as
\begin{eqnarray}
S_{qq}(b^+,b^-,b_\perp,\mu) &= \text{Tr} \langle 0 |\bar{T} [ S_n^\dagger S_{\bar{n}} ] (b^+,b^-,b_\perp) \> T [ S_{\bar{n}}^\dagger S_n ](0) |0\rangle .
\end{eqnarray}
The n-collinear iBF is  defined as a nucleon matrix element of collinear quark fields and Wilson lines in SCET as
\bea
\label{iBF-1}
\tilde{B}^{q}_{n}(x,t,b_\perp,\mu) &=& \frac{1}{2x\,\bn\cdot p_1} \int \frac{db^-}{4\pi} e^{\frac{i t}{2Qx}b^-}   \langle p_1| \bar{\xi}_{nq}W_n(b^-,b_\perp) \frac{\bnslash}{2}\nn \\
&\times& \delta(\bn \cdot {\cal P}- x\>\bn \cdot p_1) W^\dagger_n\xi_{nq}(0) | p_1 \rangle \, 
\eea
and the coefficients ${\cal I}_{n;q r}$ appearing in Eq.(\ref{intro-DY-2}) are given by the iBF to PDF matching equation
\bea
\label{eq:quarkmatch}
\tilde{B}_{n}^q(x, t,b_\perp,\mu) &=& \int _x^1\> \frac{dz}{z}\> \Big\{{\cal I}_{n; qq'} (\frac{x}{z}, t,b_{\perp}, \mu) \>f_{q'}(z,\mu)\nn \\
&+& {\cal I}_{n; qg} (\frac{x}{z}, t,b_{\perp}, \mu) \>f_{g}(z,\mu) \Big\}.
\eea
Analogous equations exist for the  $\bn$-collinear functions. 

As can be seen from Eq.(\ref{iBF-1}), the iBF corresponds to a fully unintegrated PDF. This is in contrast with the TMDPDF which has been
extensively studied in the context of transverse momentum distributions. The TMDPDF is typically defined with reference to  an external regulator to regulate
spurious rapidity divergences that arise in perturbative calculations. The independence of the cross-section from this external regulator gives rise
to the Collins-Soper evolution equation which sums large logarithms. In contrast, the iBF is free of rapidity divergences. They are automatically regulated by
the additional residual light-cone momentum component ($k^+$) which appears through the definition of the convolution variable $t=Qk^+$. This
is a reflection of the fact that for a finite $p_T$ gluon recoiling against the heavy boson, the mass-shell condition $p^+p^-=p_T^2$ ensures
a non-zero residual light-cone momentum component so that no rapidity divergence occurs in the physical process.
 The use of the residual light-cone momentum 
component to regulate rapidity divergences differs from the TMDPDF approach where an external regulator is introduced instead. The iBF
is more differential than the TMDPDF and  corresponds to a fully unintegrated PDF.  A more detailed
understanding of the relationship between the TMDPDF and the iBF approaches requires further study. However, an explicit comparison of the iBF
approach with the TMDPDF or CSS formalism is possible by expanding the resummed results and comparing the logarithms generated at any fixed order
in perturbation theory. Such an explicit check of the leading and next-to-leading logs  in the iBF approach was performed in Ref.~[\refcite{Mantry:2010mk}].

\section{Numerical Results}

\begin{figure}[h]
\includegraphics[angle=90,scale=0.4]{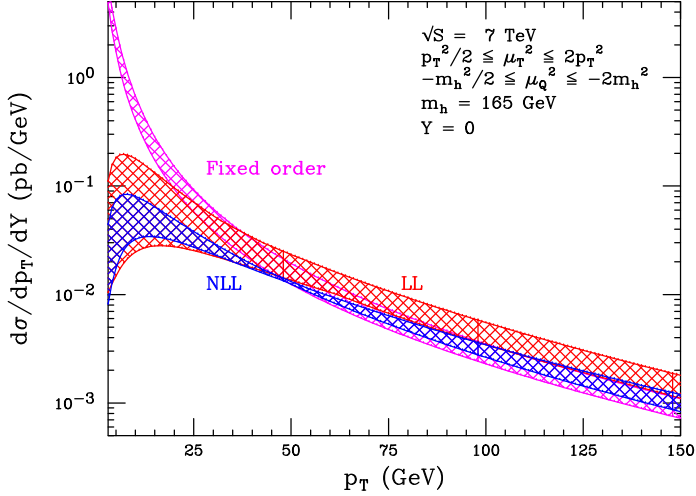}
\caption{Numerical predictions for the transverse momentum spectrum for Higgs boson production at the LHC for central rapidity.  Shown are the fixed-order result and those obtained after implementing the resummation formula of Eq.~(\ref{intro-DY}) through LL and NLL.  The bands arise from the scale variation.}
\label{higgsplot}
\end{figure}

In this section, we present numerical results for the Higgs boson and Z-boson $p_T$ distributions. We refer the reader to Refs.~[\refcite{Mantry:2009qz,Mantry:2010mk}] for analytic expressions
for the iBFs and the iSF.   In Fig.~\ref{higgsplot}, we show the fixed order, Leading-Log (LL), and Next-to-Leading Log (NLL) resummed results for Higgs production at $\sqrt{s}=7$ TeV and zero rapidity for a Higgs mass of $m_h=165$ GeV. 
The
plot is cut off at $p_T=3$ GeV so that the distribution is given entirely in terms of perturbatively calculable functions and the standard PDFs. Study of the distribution
for non-perturbative values of $p_T$ requires a model for the TMF function. We describe our modeling procedure later for the case of Z production. From Fig.~\ref{higgsplot}, we see that
the effect of the resummation is dramatic and changes the shape of the distribution. In particular, the $1/p_T^2$ singular behavior  of the fixed result is brought under
control by
resummation.

\begin{figure}
\includegraphics[angle=90,scale=0.4]{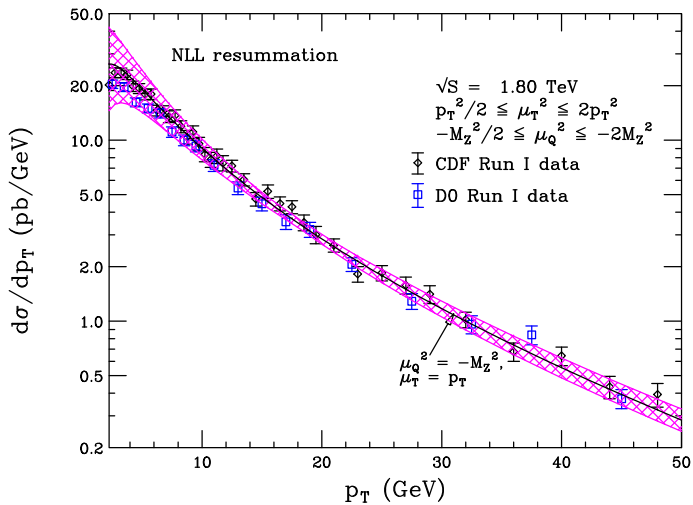}
\caption{Numerical predictions for the transverse momentum spectrum for $Z$ boson production at Tevatron Run 1, compared with data from both CDF and D0.  Shown is the resummation prediction  at NLL accuracy.  The bands arise from the scale variation, while the result for the central scale choice is shown by the solid line.  The lower limit of the plot is $p_T$= 1.75 GeV.}
\label{Zplot}
\end{figure}

Similarly, we present NLL resummation results for the Z-boson distribution for perturbative values of $p_T$ in Fig.~\ref{Zplot}. Also given is a comparison
 with Tevatron data collected by the CDF and D0 collaborations~[\refcite{Affolder:1999jh,Abbott:1999yd}]. The plot in Fig.~\ref{Zplot} is cutoff at $p_T\sim 1.75$ GeV. As seen, good agreement is found with the Tevatron data. A description of the Z-boson $p_T$ spectrum in the region $p_T \sim \Lambda_{QCD}$ requires a model for the Z-boson TMF function.  

In Ref.~[\refcite{Mantry:2010bi}], a phenomenological study of TMF models was carried out. The guiding principles for the construction of TMF models include reproducing the correct RG evolution properties and requiring that the TMF model reduces to the perturbative result as one increases $p_T$. These principles are encoded
by writing the the TMF as a convolution~[\refcite{Ligeti:2008ac,Hoang:2007vb,Fleming:2007qr,Fleming:2007xt}] of the perturbative expression of the TMF with a non-perturbative model function as
 \bea
\label{mod1}
{\cal G}^{qrs}(x_1,x_2,x_1',x_2',p_T,Y,\mu_T) &=& \int _0^\infty dp_T' \>{\cal G}^{qrs}_{\text{part.}}(x_1,x_2,x_1',x_2', p_T\sqrt{1+(p_T'/p_T)^2}, Y,\mu_T)\nn \\
&\times&\> G_{mod}(p_T', a, \Lambda), \nn \\
\eea
where $G_{mod}$ denotes a model function that peaks near $p_T' \sim \Lambda_{QCD}$ with parameters $a,\Lambda$ that can be fit to data. The scale dependence of the TMF is contained entirely in the perturbative expression ${\cal G}^{qrs}_{\text{part.}}$ of the TMF so that the required RG evolution properties are reproduced. For phenomenological analysis we parameterize $G_{mod}$ as
\bea
\label{gmod}
G_{mod}(p_T',a, \Lambda) &=& \frac{2^{3/2-a}}{\Lambda}\frac{1}{\Gamma(a-1/2)} \Bigg (\frac{p_T'^{\>2}}{\Lambda^2}\Bigg )^{a-1} \text{exp} \Big [- \frac{p_T'^{\>2}}{2\Lambda^2}\Big],
\eea
\begin{figure}
\includegraphics[height=5.5in,angle=90]{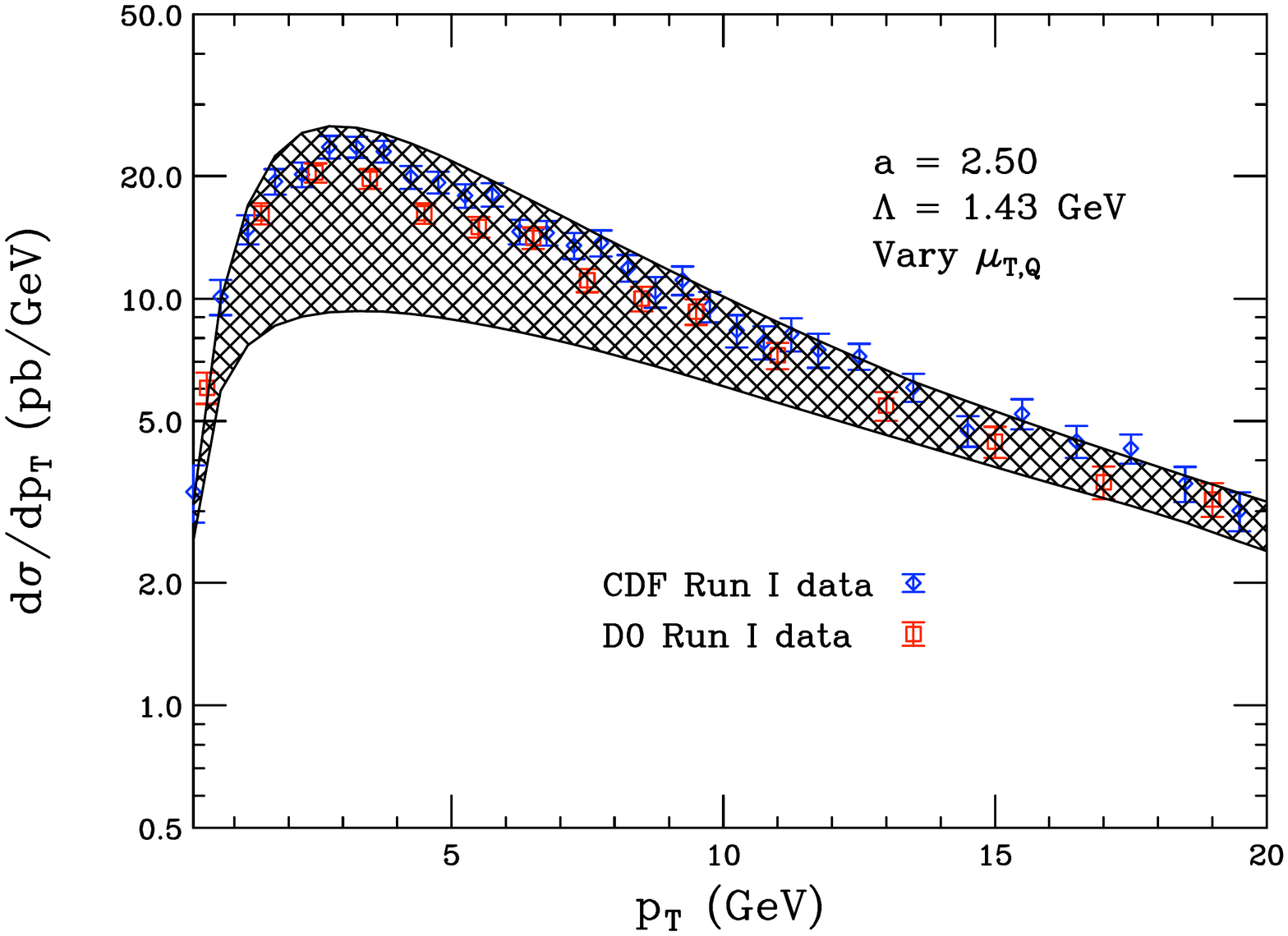}
\caption{The result for the $p_T$-spectrum of the Z-boson for the best fit parameter choices $a=2.50, \Lambda=1.43\,\text{GeV}$.  We have varied $\mu_{T,Q}$ within the range $1/2 < \xi_{T,Q} <2$ for  $\mu_T^2 = \xi_T p_T^2 + p_{Tmin}^2, \mu_Q^2=-\xi_Q M_Z^2$ for $p_{Tmin}=2 $ GeV.}
\label{fig:ZpT2}
\end{figure}
with the normalization condition
\bea
\label{norm}
\int _0^\infty dp_T' \> G_{mod}(p_T', a,  \Lambda) &=& 1.
\eea
A sensible choice for $\mu_T$ that can be applied in both the perturbative and 
non-perturbative $p_T$ regions is
\bea
\label{muTmod}
\mu_T^2 = \xi^2 \> p_T^2 + p_{Tmin}^2,
\eea
where $p_{Tmin}> 1$ GeV is a low, but still perturbative, scale and can be viewed as another parameter of the model.   $\xi$ is a scale variation parameter 
we take to be ${\cal O}(1)$. The above choice of scale for $\mu_T$ has several useful properties. As $p_T\to 0$, the scale $\mu_T\to p_{Tmin}$ 
so that ${\cal G}^{qrs}_{part}$ in Eq.~(\ref{mod1}) is still evaluated at a perturbative scale. Similarly, the running of the hard function 
$H_Z^{q}(x_1x_2Q^2,\mu_Q;\mu_T)$ will freeze at the perturbative scale $p_{Tmin}$ as $p_T\to 0$. For larger values of $p_T\gg p_{Tmin}$, 
$\mu_T \to \xi \> p_T$ so that the appropriate choice of $\mu_T \sim p_T$ in the perturbative region is recovered.
For $p_T\gg \Lambda_{QCD}$, the model TMF reduces to the expected perturbative expression up to power corrections as
\bea
\label{GOPE}
{\cal G}^{qrs}(x_1,x_2,x_1',x_2',p_T,Y,\mu_T)\Big |_{p_T\gg \Lambda_{QCD}} &=& {\cal G}^{qrs}_{\text{part.}}(x_1,x_2,x_1',x_2', p_T, Y,\mu_T) 
+ {\cal O}(\frac{\Lambda_{QCD}}{p_T}), \nn \\
\eea
so that a smooth transition between the non-perturbative and perturbative regions is achieved. In Fig.~\ref{fig:ZpT2}, we show
the Z-boson $p_T$ spectrum including the non-perturbative $p_T$ region. The values of the parameters $a=2.50, \Lambda=1.43$ GeV in $G_{mod}$ used in Fig.~\ref{fig:ZpT2}, are obtained from a best fit to the CDF and D0 data for $p_T < 10$ GeV. We see that the form of the model TMF is flexible enough to describe data over the entire range of $p_T$.

\section{Conclusions}

We have presented a new factorization theorem for the low transverse momentum distributions of electroweak gauge and Higgs bosons using the SCET.
The factorization theorem is formulated in terms of Impact-parameter Beam Functions (iBFs) which correspond to fully unintegrated PDFs. We present results of NLL resummation for Higgs and Z-boson distributions. Good agreement is found with data collected by the CDF and D0 collaborations at the Tevatron. More work is in progress  to achieve NNLL results; recently in Ref.~[\refcite{Li:2011zp}], NNLO results for the soft function of the Z-boson $p_T$ distribution were obtained. We look forward to the further development of our formalism and applications to phenomenology.

\section*{Acknowledgments}
This work is supported in part by the U.S. Department of Energy, Division of High Energy Physics, under contract DE-AC02-06CH11357 and the 
grants  DE-FG02-95ER40896 and DE-FG02-08ER4153, and by Northwestern University.

\bibliographystyle{h-physrev3.bst}
\bibliography{tmf}

\end{document}